\documentclass[preprint2,numberedappendix]{emulateapj-rtx4}

\usepackage{graphicx}
\usepackage{ulem}
\newcommand\Rm{{\rm Rm}}

\newcommand\Ru{{\rm Re}}

\begin{document}
\title{
Joint inverse cascade of magnetic energy and magnetic helicity in MHD turbulence
}
\author{R.Stepanov$^{1,2}$, P.Frick$^{1}$, I.Mizeva$^{1}$}
\affil{
$^1$Institute of Continuous Media Mechanics, Korolyov str.\ 1,
614013 Perm, Russia\\
$^2$Perm National Research Polytechnic University, Komsomolskii Av. 29, 614990 Perm, Russia
}

\label{firstpage}

\begin{abstract}

We show that oppositely directed fluxes of energy and magnetic helicity coexist in the inertial
range in fully developed magnetohydrodynamic (MHD) turbulence with small-scale sources of magnetic
helicity. Using a helical shell model of MHD turbulence, we study the high Reynolds number
magnetohydrodynamic turbulence for helicity injection at a scale that is  much smaller than the
scale of energy injection. In a short range of scales larger than the forcing scale of magnetic
helicity,  a bottleneck-like effect appears, which results in a local reduction of the spectral
slope. The slope changes in a domain with a high level of relative magnetic helicity, which
determines that part of the magnetic energy related to the helical modes at a given scale. If the
relative helicity approaches unity, the spectral slope tends to $-3/2$. We show that this energy
pileup is caused by an inverse cascade of magnetic energy associated with the magnetic helicity.
This negative energy flux is the contribution of the pure magnetic-to-magnetic  energy transfer,
which vanishes in the non-helical limit. In the context of astrophysical dynamos, our results
indicate that a large-scale dynamo can be affected by the magnetic helicity generated at small
scales. { The kinetic helicity, in particular, is not involved in the process at all. } An
interesting finding is that an inverse cascade of magnetic energy can be provided by a small-scale
source of magnetic helicity fluctuations without a mean injection of magnetic helicity.

\end{abstract}


\section{Introduction}

Magnetohydrodynamic (MHD) turbulence is an important part of astrophysical processes, which gives
rise to global cosmic magnetic fields. Over the last few decades, the peculiarities of MHD
turbulence have attracted the interest of researchers in astrophysics and fluid dynamics,
stimulating numerous theoretical and numerical studies. Recently,  significant attention has been
paid to the role of magnetic helicity in fully developed MHD turbulence. Magnetic helicity,
together with the energy and cross-helicity, is one of the three integrals of motion in ideal MHD,
but, compared to energy, the dimensions of helicity have an extra length unit, so magnetic helicity
is prone to the inverse cascade and condensation at the largest available scales
\citep{1975JFM....68..769F,1993noma.book.....B}. {The so-called catastrophic quenching problem
\citep{1992ApJ...393..165V,2000ApJ...534..984B} is an example of the importance of taking into
account the conservation law of magnetic helicity. In consequence, the growth rate of a large-scale
magnetic field under large magnetic Reynolds number is predicted to be much less than is required
for cosmic dynamos. The transparent boundary and effective transport of the magnetic helicity are
usually considered to get rid of the disagreement between theory and astrophysical observations
\citep{2005PhR...417....1B}.} The construction of the corresponding equations is still the subject
of discussion \citep{2012ApJ...748...51H}. Verification of such models might be possible with
expected observations of magnetic helicity in real cosmic fields: estimations of magnetic helicity
in solar active regions have been performed recently
\citep{2014ApJ...784L..45Z,2014ApJ...785...13L}, and there is a possibility that magnetic helicity
will be detected in the interstellar medium \citep{2014ApJ...786...91B}.

The separation of magnetic helicity into the  large-scale and small-scale terms is a standard
approach in mean-field models of large-scale dynamos, where the effect of turbulence is taken into
account through the components of the effective electromotive force \citep{Krause1980}. This
separation leads to the problem of correctly estimating of the influx of the magnetic helicity,
generated at small (subgrid) scales, into the large scales described by the mean-field equations
\citep{Frick2006}. { We stress that only the numerical simulations resolving the whole range of
scales seem to be capable of highlighting all large-scale dynamo mechanisms.}

There are a few reliable results  on the spectral transfer of magnetic helicity. The role of
magnetic helicity has been studied relatively well in free decaying MHD turbulence. Under free
decay, the magnetic helicity of the initial field draws off some of the magnetic field energy in
the largest scales.  As a result, the energy dissipation follows different scenarios of decay,
determined by the initial distribution of the magnetic helicity
\citep{Frick2010,2014arXiv1404.2238B}. Direct numerical simulations (DNS) of statistically
stationary turbulence with a substantially helical magnetic field are  complicated because they
require adequate separation of the forcing scale and dissipation scale for the energy and magnetic
helicity. An attempt at this kind of simulations was performed by \citet{Alexakis2006}, who showed
that the inverse cascade of the magnetic energy and helicity is provided by local interactions
during turbulence development and by non-local interactions in the saturated state. However, in
this model, the spectral fluxes of magnetic helicity and energy were not separated. The significant
direct cascade of magnetic helicity obtained can be explained by the proximity of the dissipation
scale to the forcing scale.

Here, we try to highlight  the role of magnetic helicity by separating its source from the source
of energy. We consider MHD turbulence that is stationary forced at the largest scale, with a source
of magnetic helicity that is localized at a scale inside the { pronounced} inertial range. In our
research, we focus on the possibility of a simultaneous direct cascade of energy  and {\it
oncoming} inverse cascade of magnetic helicity, and we examine the influence of the magnetic
helicity on the standard Kolmogorov energy cascade. { Adding an {\it ad hoc} dissipation term at
large scale helps to achieve a statistically stationary state. So we deal with the dynamo processes
at large Reynolds numbers involving MHD turbulence effects which are of obvious astrophysical
interest.}

% We suggest the model with an isolated
%effect of the magnetic helicity which let us to distinguish the key generation mechanism of the
%large-scale magnetic field. It has an impact on understanding of astrophysical dynamos where
%Reynolds numbers are typically very large.}

\section{Model of MHD turbulence}

Studying fully developed turbulent flows  demands numerical
simulations that clearly resolve the forcing, inertial and
dissipation scales. Under a condition of sufficient scale
separation, one can produce the universal behavior of turbulent
transport of three ideal quadratic invariants known in 3D
incompressible magnetohydrodynamics: the total energy $E=\langle
|{\bf u}|^2+|{\bf b}|^2\rangle/2$, the cross-helicity $H^c=
\langle{\bf u}\cdot {\bf b}\rangle$ and the magnetic helicity
$H^b=\langle{\bf a}\cdot {\bf b}\rangle$, where ${\bf u}$ is the
velocity field, ${\bf b}$ is the magnetic field, ${\bf a}$ is the
vector potential (${\bf b}={\rm \nabla}\times {\bf a}$) and
$\langle . \rangle$ means volume averaging. However, even recent
direct numerical simulations using billions of grid points hardly
provided an inertial range of scales wider than one decade
\citep{2007PhRvL..99y4502M}. This is the reason we turn to the
shell models of turbulence. These models cannot take into account
the spatial complexity of turbulent flows but reflect such
properties of real MHD turbulence as spectral distributions,
intermittency and chaotic reversals of large-scale modes (see,
e.g. \citet{2013PhR...523....1P}). Furthermore, these models
produce an extended inertial range and accurate dissipation rate
using realistic values for the governing parameters (high kinetic
and magnetic Reynolds numbers, low or high magnetic Prandtl
number).

Shell models are low-dimensional dynamic systems that are derived
from the original MHD equations by a drastic reduction of the
number of variables. These models describe the dynamics and
spectral distributions of fully developed MHD turbulence through a
set of complex variables $U_n$, $B_n$, which characterize the
kinetic energy $E_n^u=|U_n|^2/2$ and magnetic energy
$E_n^b=|B_n|^2/2$ of pulsations in the wave number range
$k_n<|{\bf k}|<k_{n+1}$ (called the shell $n$), where
$k_n=\lambda^n$ ($\lambda$ is the shell width in a logarithmic
scale, typically chosen to be equal 1.618). Model equations are
\begin{eqnarray}
d_t U_n={W_n}({\bf U},{\bf U})-{W_n}({\bf B},{\bf B})-k_n^2\frac{U_n}{\Ru}+F_n, \phantom{XX}\label{eq} \\
d_t B_n={W_n}({\bf U},{\bf B})-{W_n}({\bf B},{\bf U})-k_n^2\frac{B_n}{\Rm}+G_n-D_n, \nonumber
\end{eqnarray}
where $\Ru$ and $\Rm$ are the kinetic and magnetic  Reynolds
numbers, ${\bf U}=(U_0,...U_{N-1})$ and ${\bf B}=(B_0,...B_{N-1})$
are vectors in space $\mathbb{C}^N$, and $N$  is the total number
of shells.  The structure of equations (\ref{eq}) mimics the
original MHD equations -- the bilinear form ${\bf W}({\bf X},{\bf
Y})$ is like ${\bf X}\cdot\nabla{\bf Y}$ and terms $F_n$, $G_n$
and $D_n$ specify external forces acting in the shell $n$. We use
the $W_n({\bf X},{\bf Y})$ suggested by \citet{Mizeva2009}, which
can be rewritten in a general form as
\begin{eqnarray}
W_n({\bf X},{\bf Y})=ik_n[(X_{n-1}Y_{n-1}+X_{n-1}^*Y_{n-1}^*)-\lambda
X_n^*Y_{n+1}^*\nonumber\\
-\frac{\lambda^2}{2}(X_n Y_{n+1}+X_{n+1}Y_n+X_nY_{n+1}^*+X^*_{n+1}Y_n)\nonumber\\
-\frac{\lambda}{2}(X_{n-1}^*Y_{n-1}-X_{n-1}Y_{n-1}^*)
+\lambda X^*Y_{n+1}] \nonumber
\\ -i k_n \lambda^{-5/2}[\frac{1}{2}(X_{n-1}Y_n+X_nY_{n-1})+\lambda
X^*_nY_{n-1}^*\nonumber\\
-\lambda^2(X_{n+1}Y_{n+1}+X_{n+1}^*Y_{n+1}^*)+\frac{1}{2}(X_nY_{n-1}^*+X_{n-1}^*Y_{n})
\nonumber\\
 -\lambda X^*_nY_{n-1}+\frac{\lambda}{2}(X_{n+1}^*Y_{n+1}-X_{n+1}Y_{n+1}^*)].\phantom{XX}
\end{eqnarray}
In the non-dissipative limit, equations (\ref{eq}) conserve the
total energy $E=\sum(E^u_n+E^b_n)$, the cross-helicity
$H^c=\sum(U_nB_n^*+B_nU_n^*)/2$ and the magnetic helicity $H^b=
\sum \imath k_n^{-1}((B_n^*)^2-B_n^2)/2$. For a comprehensive
review of MHD shell models, refer to \citet{2013PhR...523....1P}.

The key point of our modelling is a particular forcing design to
create stationary cascades. We excite the turbulence in the
classical way by injecting kinetic energy at the largest scale.
Namely, we take $F_n$ at $n=0$ ($k=1$) only, in the form
\begin{equation}
F_0=I_f \exp(\imath\phi) \label{Eq.F0}
\end{equation}
where $I_f$ is constant and $\phi$ is a random phase that remains
constant during each time interval $t_c$. Then, the time-averaged
energy injection rate becomes $\varepsilon=I_f^2 t_c$, and the
injection rates of the  kinetic helicity and cross-helicity vanish
under the condition that $t_c$ is shorter than the large-scale
turnover time.

The magnetic forces  $G_n$ and $D_n$ are introduced to simulate
the source and sink of magnetic  helicity, acting at shells $n_g$
and $n_d$, respectively. For magnetic helicity injection, the
force is taken as
\begin{equation}
G_n=\frac{\imath I_g k_n B_n(B_n^2+(B_n^*)^2)}{16 (E_n^b)^2}
\label{Eq.F1}
\end{equation}
with $n=n_g$. Then, the injection rate for magnetic helicity is
$\chi=I_g (1-(H_n^r)^2)$, where $H_n^r=k_n H_n^b/(2E_n^b)$  is the
relative magnetic helicity. We note here that the force
(\ref{Eq.F1}) does not change the magnetic energy and becomes zero
for marginal values $H^r_n=\pm1$.

To produce an stationary inverse cascade of magnetic helicity and avoid its accumulation at the largest scale, we introduce a
large-scale sink of the magnetic helicity as an additional
dissipation
\begin{equation}
D_n = \frac{I_d k_n^2 B_n (H_n^b)^2}{8 (E_n^b)^3}, \label{Eq.F3}
\end{equation}
with $n=n_d$. This gives magnetic energy dissipation with a  rate $\varepsilon_d= I_d (H_n^r)^2$
and magnetic helicity dissipation with a rate $\chi_d=2 I_d k_n^{-1} (H_n^r)^3$. In the non-helical
case ($H_n^r\to 0$), this dissipation tends to zero. { The force (\ref{Eq.F3}) imitate  in the
shell model the real open boundary condition typical for astrophysical objects with cosmic magnetic
field.}

\section{Results}

\begin{figure}
\vspace{0.6cm} \hspace{8cm}(a) \par \vspace{6cm} \hspace{8cm}(b) \par \vspace{-7.3cm}
\includegraphics[width=0.49\textwidth]{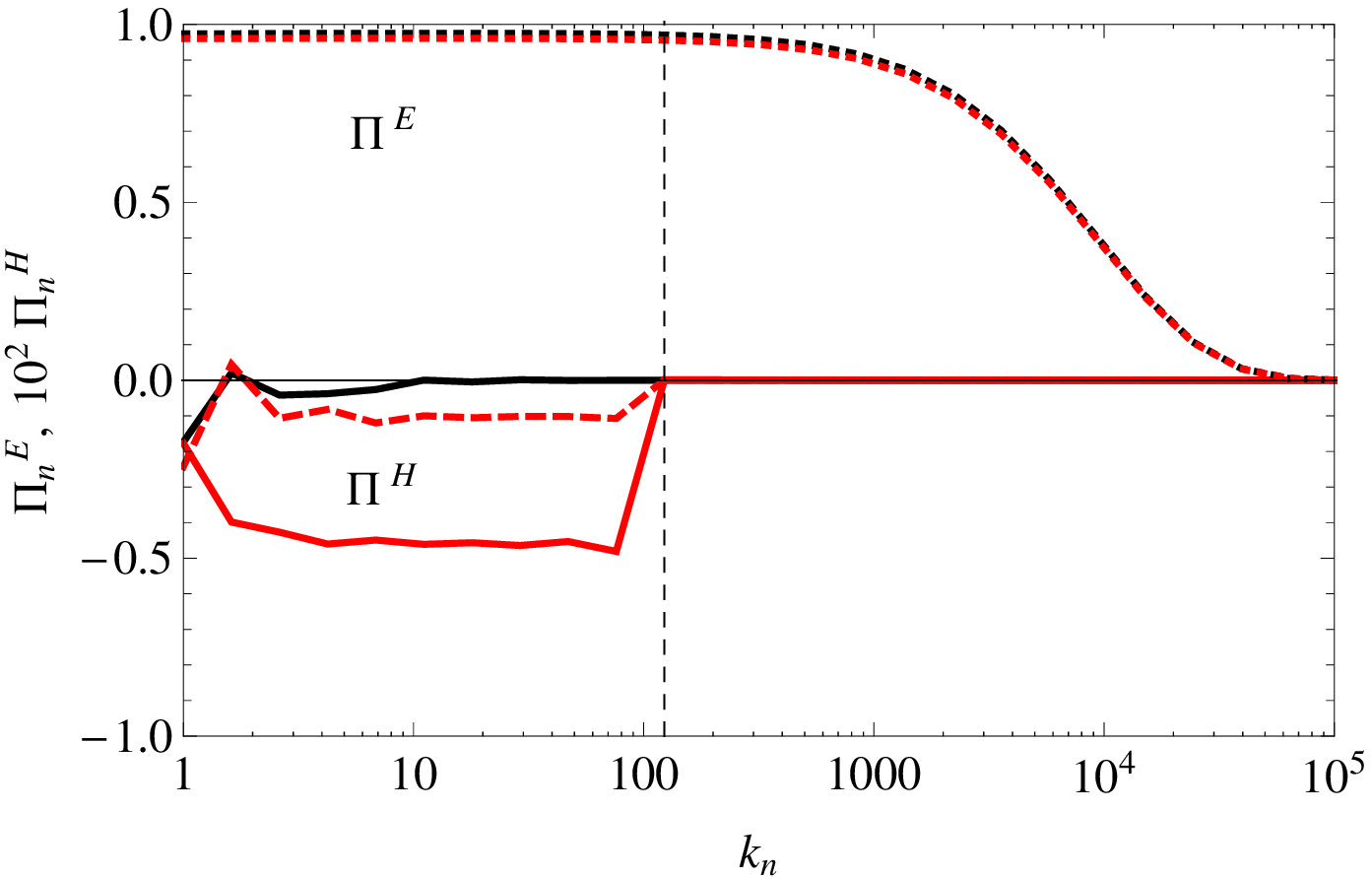}
\includegraphics[width=0.49\textwidth]{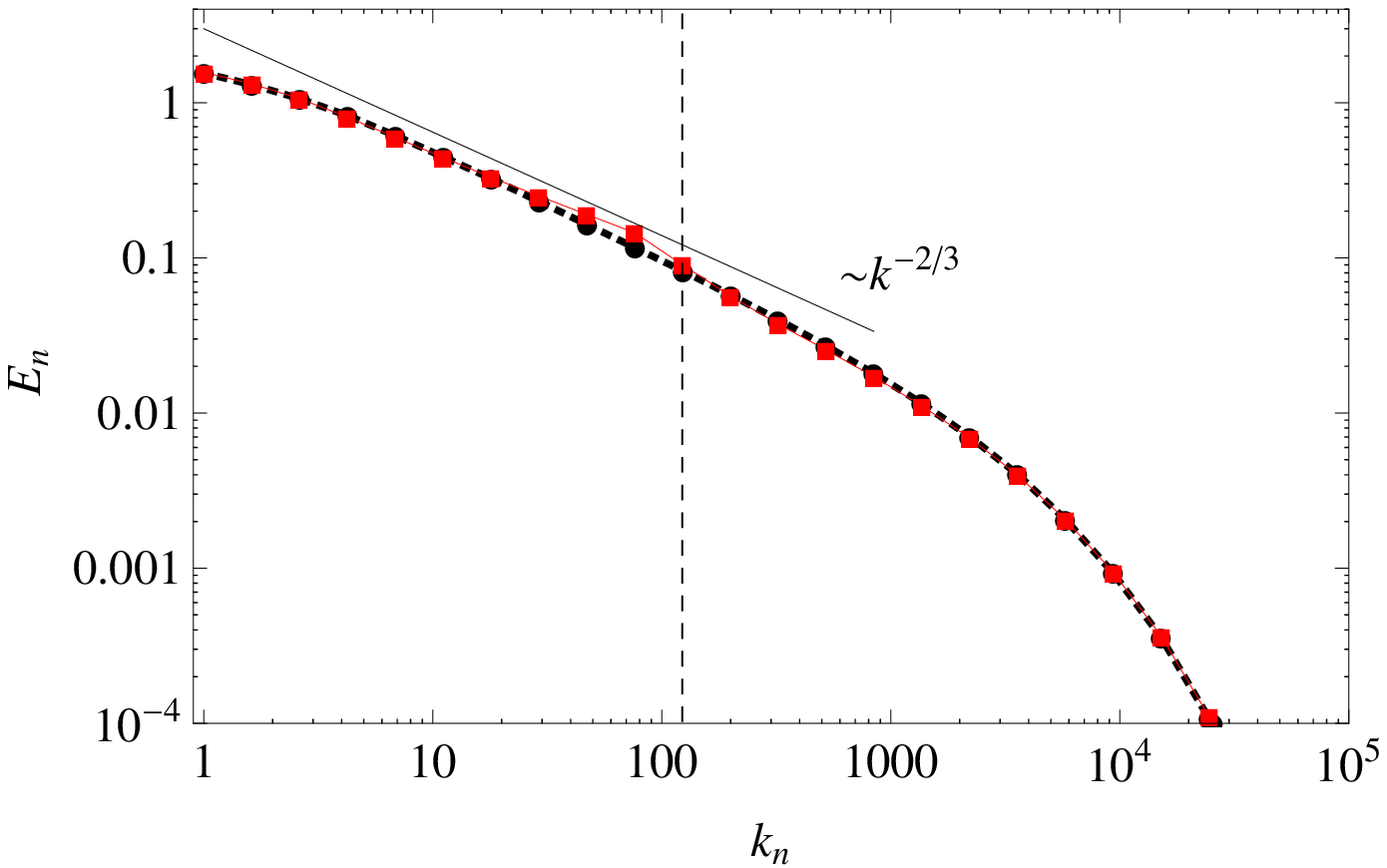}
\caption {Spectral fluxes (a) and spectra (b)  of the total energy
for stationary forced MHD turbulence. Black lines correspond to
the non-helical case. Red lines show the spectrum and fluxes for
magnetic helicity injection at wave number  $k_g=122$ indicated by a
vertical dashed line. The spectral fluxes of the total energy
$\Pi^E_n$ are shown by a dotted lines. The spectral fluxes of the
magnetic helicity $\Pi^H_n$ multiplied by a factor $10^2$ are shown
for three injection rates: $I_g=0$ (black solid), $I_g=0.05$ (red dashed)
and $I_g=0.5$ (red solid). The thin straight line corresponds to the
Kolmogorov's slope.} \label{spec1}
\end{figure}

Our reference case of simulations is stationary forced MHD
turbulence  without injection of magnetic helicity. We numerically
evolve Equations~(\ref{eq}) for $\Ru=\Rm=10^6$ and use amplitude
$I_f=10$ and $t_c=0.01$ for the force (\ref{Eq.F0}), which
provides the energy injection rate $\varepsilon=1$ at the scale $k=1$.
The corresponding spectrum and spectral flux of the total energy
are shown in Figure~\ref{spec1}. All curves corresponding to this
non-helical case are shown in black.  A Kolmogorov's spectral law
$E\sim k^{-5/3}$ (which corresponds to $E_n\sim k_n^{-2/3}$) and a
flat spectral flux extend for about three decades. The shell models gain an advantage over DNS with this considerable separation of forcing and dissipation scales.
Note, that all statistical quantities shown in our figures are calculated by averaging over 32 numerical realizations, performed for similar initial conditions each for
a period of $10^3$ large-scale turnover times.

Next we consider a helical case with an injection of magnetic helicity within the inertial range,
namely at $k_g=122$ ($n_g=10$), with a constant mean injection rate $\chi>0$ provided by the force
(\ref{Eq.F1}). The injected magnetic helicity is transferred toward scales larger  than $k_g$. This
means that the magnetic helicity spectral flux $\Pi^H_n$ is negative. To achieve the stationary
state of turbulence, we remove the magnetic helicity at $k=1$ ($n_d=0$) using force (\ref{Eq.F3}) {
with $I_d=1$.  This value of $I_d$ is sufficient to keep balance of the injection and dissipation
rates of magnetic helicity $\chi=\chi_d$ for any $I_g$.}  Figure~\ref{spec1}(a) shows a stable
inverse cascade of magnetic helicity with almost constant spectral flux $\Pi^H_n$ for different
values of $I_g$, which do not noticeably influence the direct cascade of energy characterized by
positive energy flux $\Pi^E_n$. Figure~\ref{spec1}(b) shows that the helicity injection does not
change the energy spectrum, except for a small bump near $k_g$.

\begin{figure}
\vspace{0.6cm} \hspace{8cm}(a) \par \vspace{6cm} \hspace{8cm}(b) \par \vspace{-7.3cm}
\includegraphics[width=0.49\textwidth]{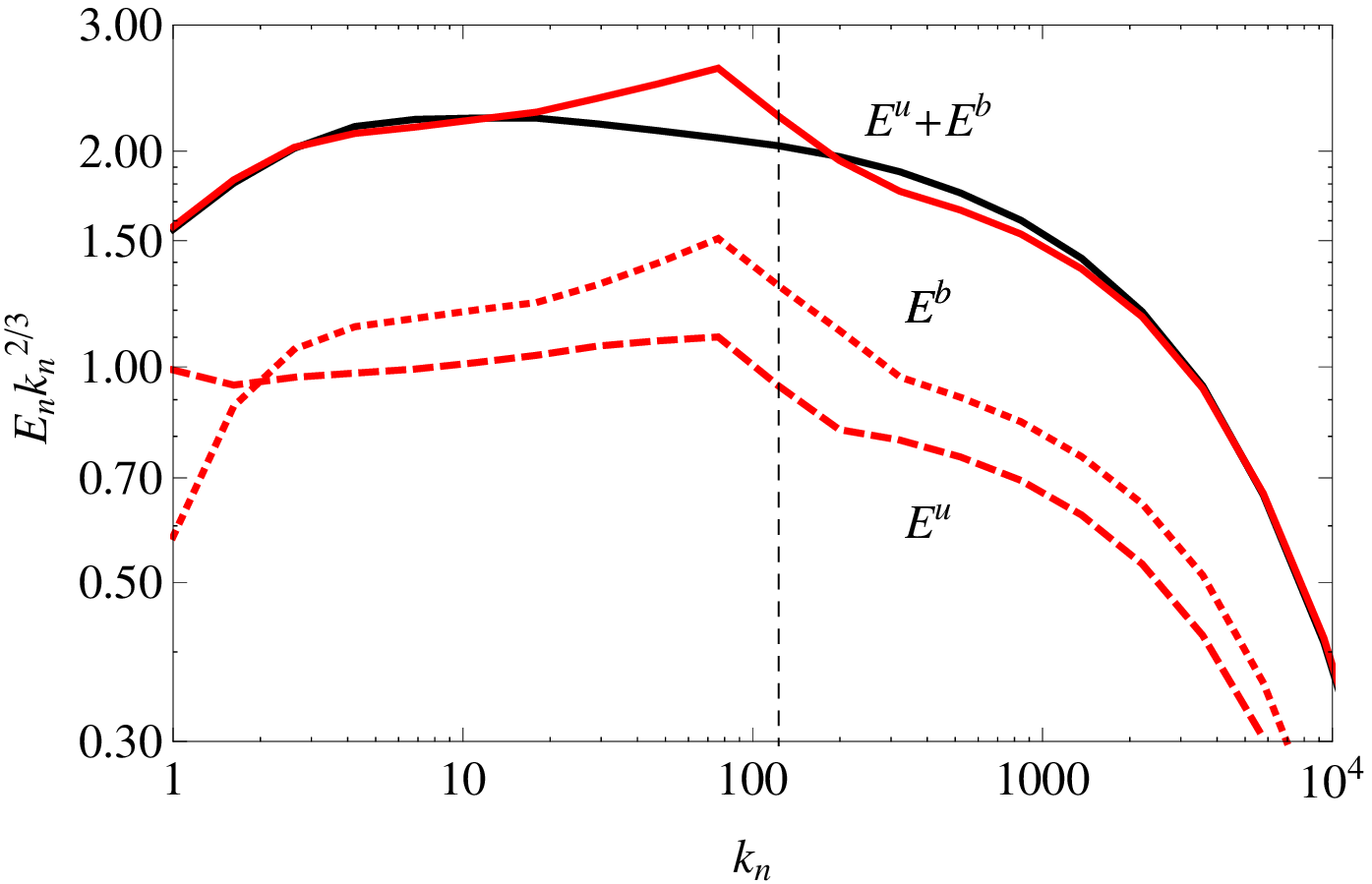}
\includegraphics[width=0.49\textwidth]{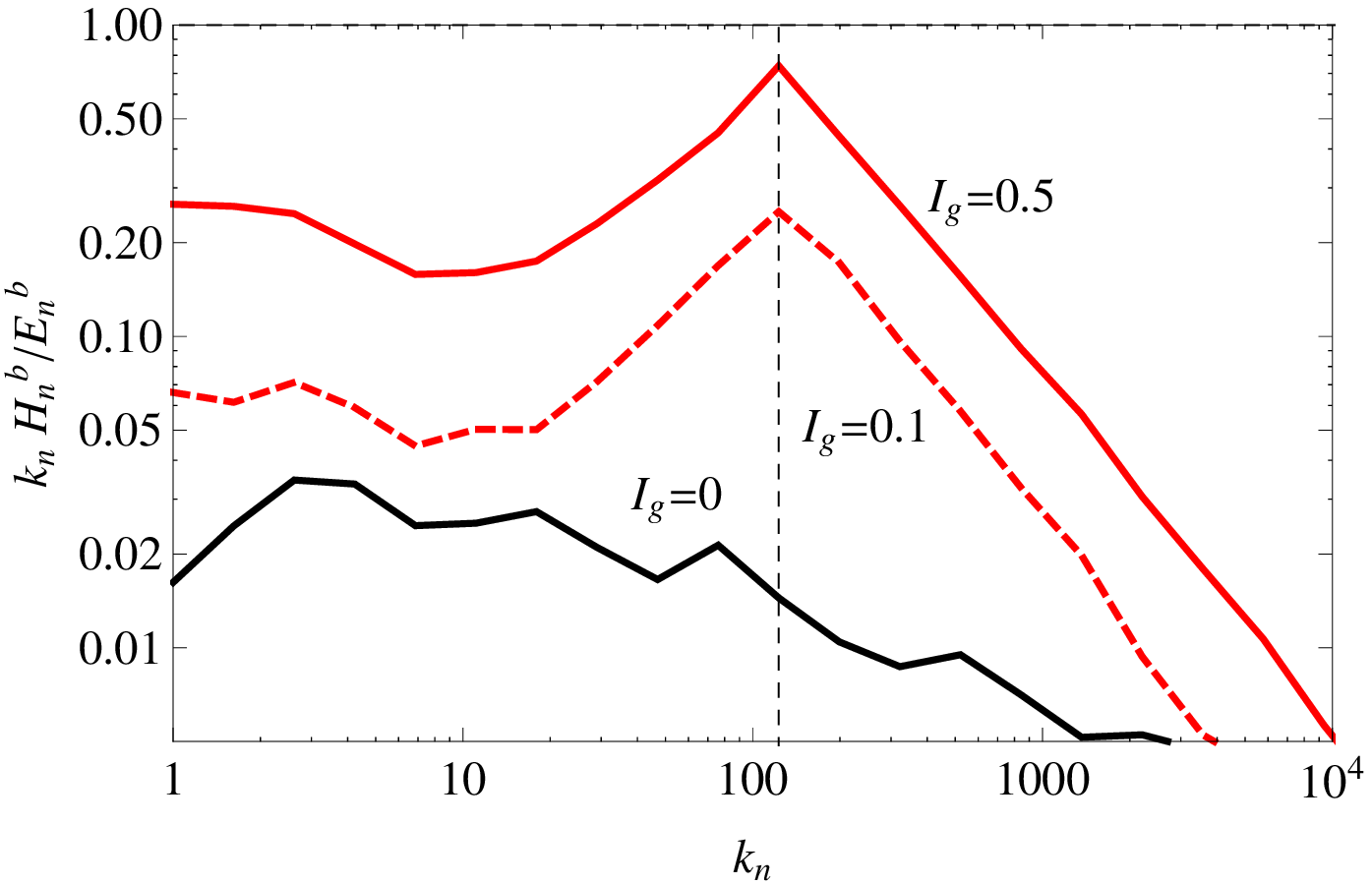}
\caption {(a) Compensated spectra of energy (kinetic, magnetic and
total) together with the spectrum of total energy for the
non-helical case (black line). (b) Spectra of relative magnetic
helicity $H^r_n$ for $I_g=0$ (black line), $I_g=0.1$ (red dashed line)
and $I_g=0.5$ (red solid line).} \label{spec1a}
\end{figure}

To emphasis the bump, we present in Figure~\ref{spec1a}(a) the
compensated spectra of the total, kinetic and magnetic energies
separately for the helical case. The non-helical spectrum is close
to horizontal, which corresponds to Kolmogorov's power law with
exponent $-5/3$. Increasing $I_g$ leads to an increase of
$\Pi^H_n$ and to the growth of relative magnetic
helicity $H^r_n=kH^b_n/2E^b_n$ over the whole spectrum, as is shown in
Figure~\ref{spec1a}(b). However, a more intensive injection than
$I_g=0.5$ does not change the situation because  $H^r_n$ reaches
the limit equal to unity and the forcing is saturated. In this
saturated state, the spectral slope at wave numbers smaller than
$k_g$  tends to the $-3/2$ power law. Recall that there is energy
injection at $k_g$ and the corresponding scale is rather far from
the energy forcing scale and the dissipation scale. The physics
behind this bump should be explained by local distortion caused by
pure magnetic helicity injection.

A kind of pileup of the  energy spectrum in the inertial range is
known for conventional developed turbulence as a result of the
bottleneck phenomenon \citep{1994PhFl....6.1411F}. In isotropic
fully  developed hydrodynamic turbulence, the bottleneck effect is
caused by edge effects, related to the transition from inertial to
diffusive scales. The viscous suppression of small-scale modes
removes some triads from non-linear interaction and makes the
spectral energy transfer less efficient, which leads to a pileup
of energy at the end of the inertial range of scales. Note that in
shell models this effect could be reproduced by using a non-local
model that includes interactions of remote shells
\citep{Plunian2007}.

\begin{figure}
\vspace{0.6cm} \hspace{2cm}(a) \par \vspace{6cm} \hspace{2cm}(b) \par \vspace{-7.3cm}
\includegraphics[width=0.49\textwidth]{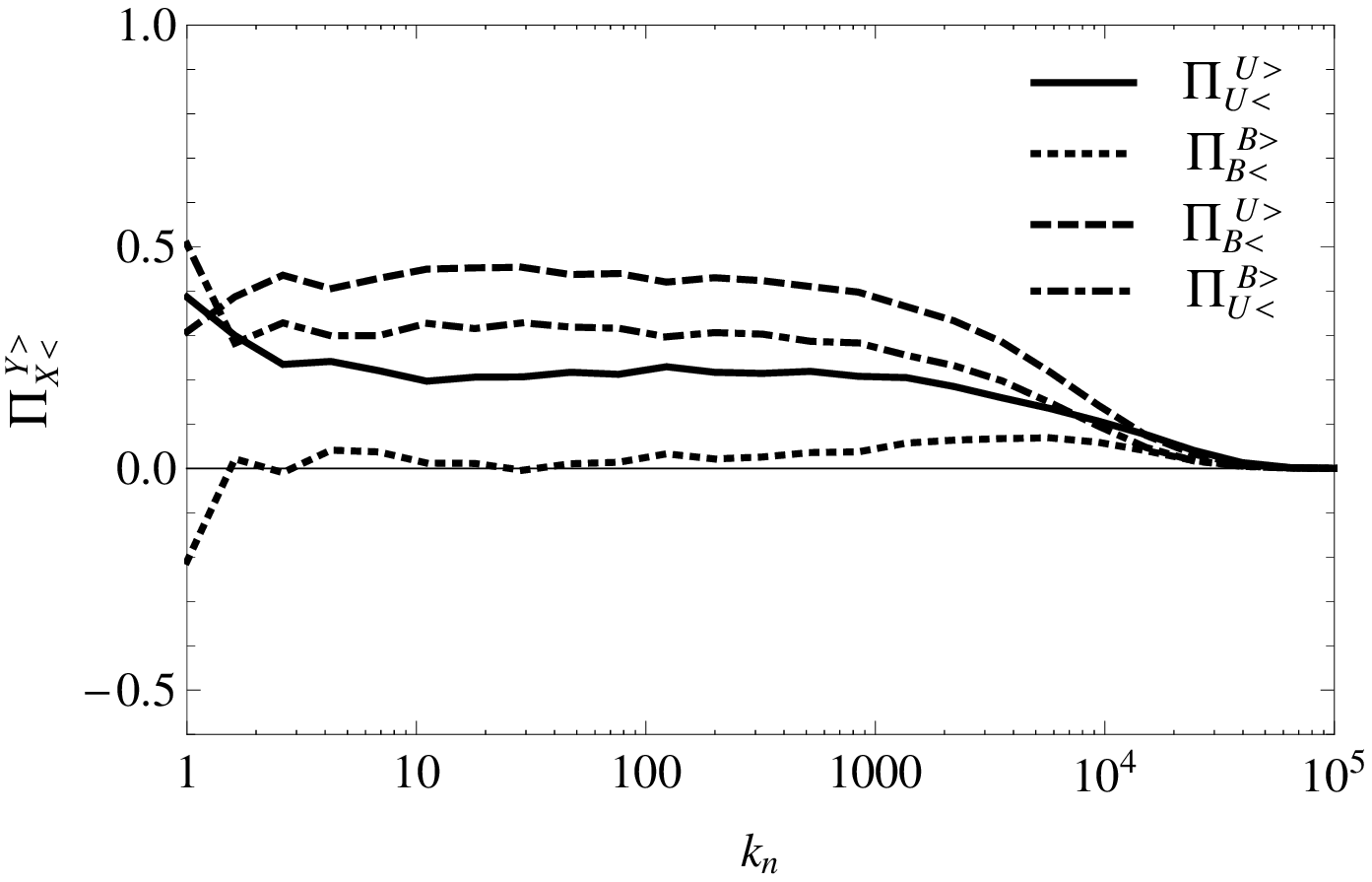}
\includegraphics[width=0.49\textwidth]{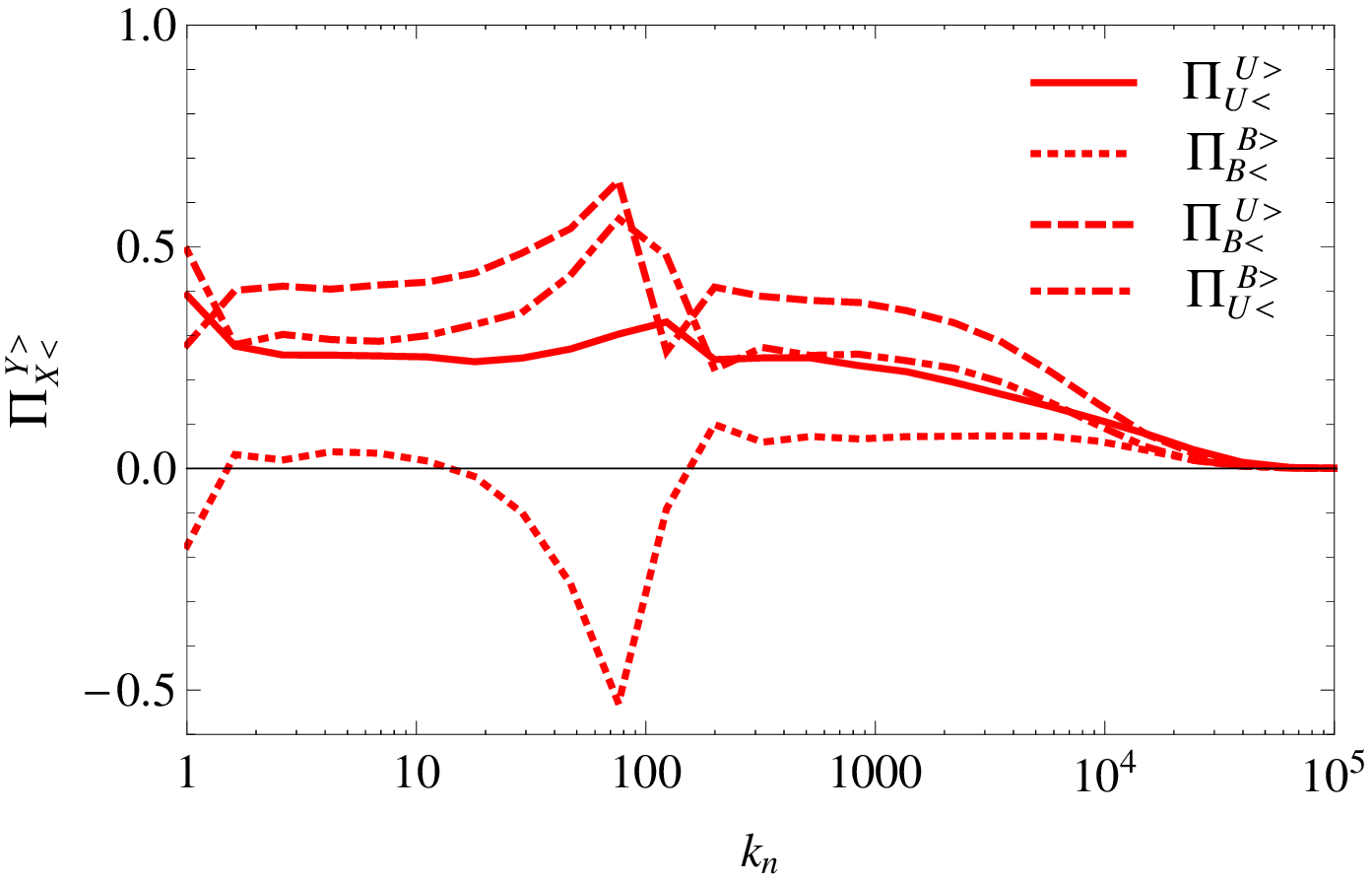}
\caption {Four components of the energy flux $\Pi_{U^<}^{U^>}$,
$\Pi_{U^<}^{B^>}$, $\Pi_{B^<}^{U^>}$ and $\Pi_{B^<}^{B^>}$ vs the
wave number $k_n$ for the non-helical case $I_g=0$ (a) and helical
case  $I_g=0.5$ (b).} \label{4fluxes}
\end{figure}

We suggest that the effect of the magnetic helicity on the energy
spectrum could be clarified by considering energy transfers
between kinetic and magnetic field modes of different scales.
Shell-to-shell energy transfer in magnetohydrodynamics has been
suggested and considered in frame of DNS by \citet{Alexakis2005}.
An analogous formalism had been derived in terms of shell models
\citep{Plunian2007}. Instead of shell-to-shell transfers, we
prefer to use the spectral fluxes as defined by
\citet{Lessinnes2009}:
\begin{eqnarray}
    \Pi_{U^<}^{U^>}=-{\bf W}({\bf U},{\bf U})\cdot {\bf U}_n^<, \nonumber\\
    \Pi_{U^<}^{B^>}=-{\bf W}({\bf B},{\bf B})\cdot {\bf U}_n^<, \label{fluxes}\\
    \Pi_{B^<}^{B^>}={\bf W}({\bf U},{\bf B}_n^>)\cdot  {\bf B}_n^<, \nonumber\\
    \Pi_{B^<}^{U^>}=-{\bf W}({\bf B},{\bf U}_n^>)\cdot {\bf B}_n^<, \nonumber
\end{eqnarray}
where each  $\Pi_{X^<}^{Y^>}$ denotes the spectral energy flux
from ${\bf X}_n^<=(X_1,X_2,... X_n,0,\dots)$ to ${\bf
Y}_n^>=(0,...Y_{n+1},Y_{n+2},...)$. In the flux notation the
subscripts $n$ has been dropped for convenience, e.g.
$\Pi_{X^<}^{Y^>}$ is a function of $n$ and  must be understood as
$\Pi_{X_n^<}^{Y_n^>}$.  Definitions (\ref{fluxes}) satisfy the
total energy conservation condition so that
$$\Pi^E=\Pi_{U^<}^{U^>}+\Pi_{U^<}^{B^>}+\Pi_{B^<}^{B^>}+\Pi_{B^<}^{U^>}.$$
Figure~\ref{4fluxes}  shows these four components of the energy
flux (\ref{fluxes}) for the non-helical and helical cases. For
non-helical turbulence (see Figure~\ref{4fluxes}(a)) the term
$\Pi_{B^<}^{B^>}$ is negligible with respect to other three, which
are constant over the inertial range. Injection of the magnetic
helicity results in the negative flux $\Pi_{B^<}^{B^>}$ (see
Figure~\ref{4fluxes}(b)). $\Pi_{B^<}^{B^>}$ has a minimum near
$k_g$ and scales as $\Pi_{B^<}^{B^>} \sim \Pi^H k$ for $k\leq
k_g$. Thus the flux of magnetic helicity is necessary associated
with the inverse flux of magnetic energy, that is described by
$\Pi_{B^<}^{B^>}$. Since the total energy flux through any wave
number inside the inertial range must be constant, fluxes
$\Pi_{U^<}^{U^>}$, $\Pi_{U^<}^{B^>}$ and $\Pi_{B^<}^{U^>}$
compensate for the drop caused by $\Pi_{B^<}^{B^>}$. One can see
the corresponding growth of these fluxes at $k_g$ in
Figure~\ref{4fluxes}(b). This growth is provided by
intensification of the kinetic and magnetic fields near $k_g$. As
a consequence, a bump forms in the energy spectrum.

Note that all the above results were obtained for positive $I_g$,
which provide an injection of positive magnetic helicity only.
Changing of the sign does not affect the results, except for the
sign of the magnetic helicity spectral flux. $\Pi^H_n$ for $k<k_g$
becomes positive corresponding to an inverse (negative) cascade of
negative helicity.

\begin{figure}
\vspace{0.6cm} \hspace{1.5cm}(a) \par \vspace{6cm} \hspace{1.5cm}(b) \par \vspace{-7.3cm}
\includegraphics[width=0.49\textwidth]{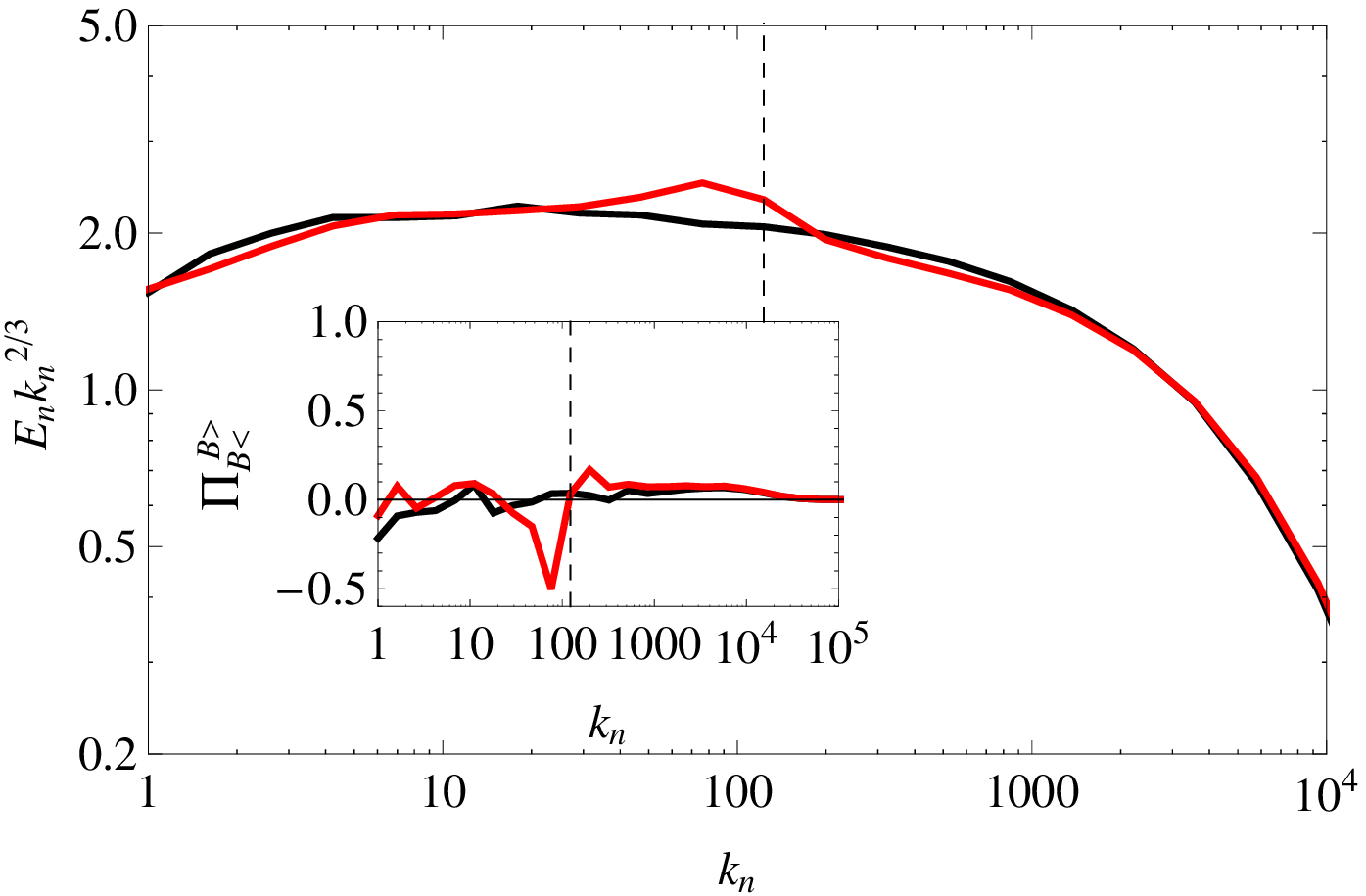}
\includegraphics[width=0.49\textwidth]{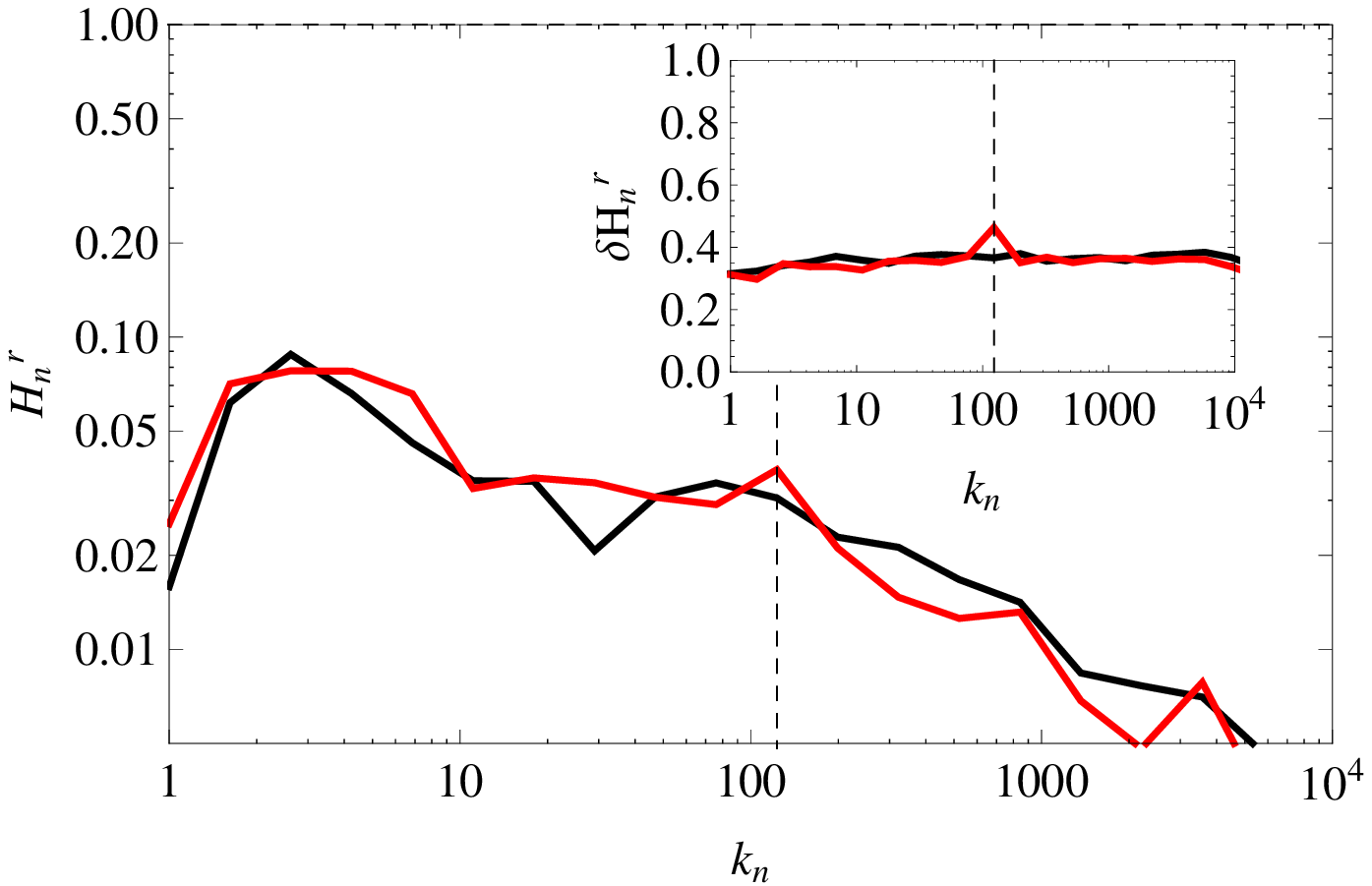}
\caption {The case with an alternating injection of magnetic
helicity:  (a) compensated spectra (the inset is the
magnetic-to-magnetic energy flux) and (b) relative magnetic
helicity (the inset shows the standard deviation of $H^r_n$.
$I_g=0$ - black curves, $I_g=1$ - red curves.} \label{fig:E1}
\end{figure}

Finally we address the question of what happens if the small-scale
source of magnetic helicity produces a fluctuating magnetic
helicity injection, being zero averaged over time. To examine this
case, we introduce a force that injects magnetic helicity in an
alternating manner. Namely the sign of the injected helicity
corresponds to the actual value of magnetic helicity. This can be
done by modifying force (\ref{Eq.F1}) via multiplying by $H^r_n$
so that
\begin{equation}
G_n=\frac{\imath I_g k_n^2
B_n H_n^b (B_n^2+(B_n^*)^2)}{(E_n^b)^3}. \label{Eq.F7}
\end{equation}
The injection rate, caused by this force, is $\chi=16 I_g
(1-(H_n^r)^2) H^r_n$. The resulting energy spectrum and magnetic
helicity flux are shown in Figure~\ref{fig:E1}(a). One can see
that the effect is similar to that obtained with the force
(\ref{Eq.F1}), which injects magnetic helicity of fixed sign. The
particularity of the result is that the averaged helicity spectrum
(see Figure~\ref{fig:E1}(b)) does not differ from the spectrum for
the non-helical case (compare black and red curves). As expected,
the force (\ref{Eq.F7}) just amplifies the amplitude of the
magnetic helicity fluctuation and increases the characteristic
time between the changes in its sign. The time during which the
magnetic helicity has the same sign becomes sufficient to initiate
the inverse cascade. Recall that injected negative helicity
cascades to large scales under a positive spectral flux. However,
the flux $\Pi_{B^<}^{B^>}$ is negative anyway. So we have found a
situation in which any characteristics of the magnetic helicity does not reveal the inverse cascade (the spectral
distribution of both magnetic helicity and its flux do not
change), while the associated flux of magnetic energy can be
detected, namely by the contribution to the energy flux, provided
by the term $\Pi_{B^<}^{B^>}$.

\section{Conclusions}

In fully developed MHD turbulence, a source of magnetic helicity at small scale provides a negative
spectral flux, which coexists with the direct energy flux in the inertial range. Near the scale of
helicity injection a bottleneck-like effect appears, which leads to a local reduction of the
spectral slope. We found that the key quantity for understanding this effect is the
magnetic-to-magnetic energy spectral flux. This flux, being negligible in the non-helical case, is
negative and clearly associated with an inverse cascade of magnetic helicity independent of the
sign of the injected helicity. The same effect can be obtained even for an alternating source of
small-scale magnetic helicity with a zero-mean injection rate. In spite of the rather special conditions in our
modelling, a similar scenario to some extend can develop in realistic situations, e.g. magnetic
helicity injection into the corona in emerging active regions \citep{2014ApJ...785...13L}.

\begin{figure}
\includegraphics[width=0.49\textwidth]{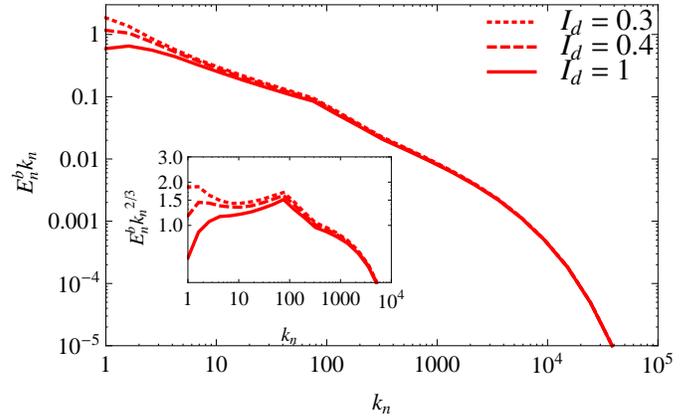}
\caption {Spectra of magnetic energy (the inset is the compensated version) the
magnetic-to-magnetic energy flux) for $I_g=0.5$ and different amplitude of the dissipative force
(\ref{Eq.F3}): $I_d=1$ (solid line), $I_d=0.4$ (dashed line) and $I_d=0.1$ ( dotted line).}
\label{fig:E4}
\end{figure}

The physical implications of the simulations presented in the paper are the following reasoning
about the large-scale dynamo mechanism. Mirror symmetry breaking of magnetic field fluctuations at
small scales initiates an inverse cascade of magnetic helicity. This leads necessarily to a
magnetic energy spectral flux from small scales to large scales, which consequently  causes the
growth of a large-scale magnetic field in the kinematic dynamo regime. The large-scale magnetic
energy becomes saturated when the inverse spectral fluxes of the magnetic energy and helicity are
compensated by corresponding outflows at the largest scale. Figure~\ref{fig:E4} shows that the
reduction of a large-scale dissipative force (\ref{Eq.F3}) results in accumulation of magnetic
energy at large scales. We believe that this qualitative scenario was implicitly assumed in earlier
studies. However, we have succeeded in demonstrating the scenario for the developed MHD turbulence,
with an extended inertial range, using a model based on local interactions of scales. We note that
mixing of local and nonlocal interactions is not avoidable even in the recent large-scale dynamo
simulations due to a lack of resolution. Actually, we have posed the problem of the realisability
of a large-scale dynamo in terms of the efficiency of mode interactions, which provide a
magnetic-to-magnetic inverse energy spectral flux and emphatically recommend this for
consideration. In addition, we have found that the magnetic helicity injection in the alternating
manner (being zero averaged in time) plays the same role.

{Apparently the mechanism considered for the magnetic energy condensation at large scale can be
interpreted as the fluctuating magnetic $\alpha$-effect, which allows the large-scale dynamo to
persist at values of the Reynolds numbers relevant for astrophysical conditions. In contrast to the
dynamo recently suggested by \citet{2013Natur.497..463T}, which was also obtained for a large-scale
magnetic field at large $\Rm$, our dynamo model does not contain any shear-like terms nor kinetic
helicity forcing. }

The result obtained is important for the theory of astrophysical dynamos, showing that a
large-scale dynamo can be affected by the magnetic helicity generated at small scales. In
particular, it means that for mean-field dynamo models one should consider the possible flux of
magnetic helicity from small (subgrid) scales. Note, that this contribution by small-scale
turbulence for a large-scale dynamo can be described by multi-scale models, which use mean-field
equations for the large scales and shell equations for the small scales
\citep{Frick2006,Nigro2011}.

\acknowledgments We thank Dmitry Sokoloff and the referee for useful suggestions that have led to
improvements of the manuscript. This work benefitted from the Russian Foundation of Basic Research
grant 14-01-96010.  Computing resources of the supercomputer URAN were provided by the Institute of
Mathematics and Mechanics UrB RAS.

%\bibliographystyle{apj}                       %% AASTeX
%\bibliography{apj-jour,ref}
%\end{document}

\end{document}